\documentclass[aps,preprint,superscriptaddress,nofootinbib,nobibnotes,a4paper,12pt,onecolumn,titlepage]{revtex4}
\usepackage{amsfonts}
\usepackage{amsmath}
\usepackage{amssymb}
\usepackage[dvips]{graphicx}

\input{tcilatex}
\begin{document}

\title{Electromagnetically induced transparency in multi-level cascade
scheme of cold rubidium atoms}
\author{J. Wang}
\affiliation{State Key Laboratory of Magnetic Resonance and Atomic and Molecular
Physics,Wuhan Institute of Physics and Mathematics,The Chinese Academy of
Sciences, Wuhan 430071,P. R. China}
\affiliation{Center for Cold Atom Physics, Chinese Academy of Sciences,Wuhan 430071,P. R.
China}
\author{L. B. Kong}
\affiliation{State Key Laboratory of Magnetic Resonance and Atomic and Molecular
Physics,Wuhan Institute of Physics and Mathematics,The Chinese Academy of
Sciences, Wuhan 430071,P. R. China}
\affiliation{Center for Cold Atom Physics, Chinese Academy of Sciences,Wuhan 430071,P. R.
China}
\author{K. J. Jiang}
\affiliation{State Key Laboratory of Magnetic Resonance and Atomic and Molecular
Physics,Wuhan Institute of Physics and Mathematics,The Chinese Academy of
Sciences, Wuhan 430071,P. R. China}
\affiliation{Center for Cold Atom Physics, Chinese Academy of Sciences,Wuhan 430071,P. R.
China}
\author{K. Li}
\affiliation{State Key Laboratory of Magnetic Resonance and Atomic and Molecular
Physics,Wuhan Institute of Physics and Mathematics,The Chinese Academy of
Sciences, Wuhan 430071,P. R. China}
\affiliation{Center for Cold Atom Physics, Chinese Academy of Sciences,Wuhan 430071,P. R.
China}
\author{X. H. Tu}
\affiliation{State Key Laboratory of Magnetic Resonance and Atomic and Molecular
Physics,Wuhan Institute of Physics and Mathematics,The Chinese Academy of
Sciences, Wuhan 430071,P. R. China}
\affiliation{Center for Cold Atom Physics, Chinese Academy of Sciences,Wuhan 430071,P. R.
China}
\author{H. W. Xiong}
\affiliation{State Key Laboratory of Magnetic Resonance and Atomic and Molecular
Physics,Wuhan Institute of Physics and Mathematics,The Chinese Academy of
Sciences, Wuhan 430071,P. R. China}
\affiliation{Center for Cold Atom Physics, Chinese Academy of Sciences,Wuhan 430071,P. R.
China}
\author{Yifu Zhu }
\affiliation{Department of Physics, Florida International University,Miami, Florida 33199}
\affiliation{Center for Cold Atom Physics, Chinese Academy of Sciences,Wuhan 430071,P. R.
China}
\author{M. S. Zhan}
\affiliation{State Key Laboratory of Magnetic Resonance and Atomic and Molecular
Physics,Wuhan Institute of Physics and Mathematics,The Chinese Academy of
Sciences, Wuhan 430071,P. R. China}
\affiliation{Center for Cold Atom Physics, Chinese Academy of Sciences,Wuhan 430071,P. R.
China}
\date{\today}

\begin{abstract}
We report an experimental investigation of electromagnetically induced
transparency in a multi-level cascade system of cold atoms. The absorption
spectral profiles of the probe light in the multi-level cascade system were
observed in cold $^{85}$Rb atoms confined in a magneto-optical trap, and the
dependence of the spectral profile on the intensity of the coupling laser
was investigated. The experimental measurements agree with the theoretical
calculations based on the density matrix equations of the rubidium cascade
system.

\end{abstract}

\maketitle

Electromagnetically induced transparency (EIT)\cite{[1]} is a quantum
interference effect that permits propagation of light through an opaque
atomic medium without attenuation, it was first proposed in 1989 \cite{[2]}
and experimentally verified in 1991 \cite{[3]}. Since then, theoretical and
experimental studies of EIT have attracted great attentions due to their
potential applications in many fields, such as low light nonlinear optics 
\cite{[4]}, quantum information \cite{[5]}, atomic frequency standard \cite%
{[6]}, and so on. Early studies were carried out with hot atoms in vapor
cells. In the hot atomic medium, the interaction time between the atoms and
the laser fields is short which leads to the transient broadening. Also, the
collisions in the hot atomic medium may severely shorten the coherence decay
time. Recently, many groups explored the EIT phenomena using the laser
cooled atoms. There are several advantages in the cold atoms \cite{[7]}.
Firstly, because of the low temperature of the cold atoms, the Doppler
broadening effect is effectively minimized, which renders it possible to
explore EIT-type nonlinear optical phenomena involving odd number of
photons. Secondly, the lower collision rates in the cold atomic sample
reduce the decoherence rate.

Early experimental studies of EIT in the cold atoms were mainly carried out
in rubidium atoms \cite{[8],[9],[10]}. Subsequently, the EIT based nonlinear
optical phenomena were studied \cite{[4],[11]}, which led to the recent
experiments on the resonant nonlinear optics at low light intensity. A very
steep slope of refractive index and the extremely low group velocity of
probe light have been obtained in the cold EIT mediums \cite{[12]}, which
have been used to demonstrate light storage and recall based on the coherent
excitation transfer between the photons and the atoms \cite{[13]}. Recently,
electromagnetically induced grating (EIG) \cite[15]{[14]} was realized in
the cold atoms. Jason et al. experimentally compared the EIT phenomena
between the hot atoms and the cold atoms \cite{[16]}, and Ahufinger et al.
compared the EIT phenomena between the cold atoms above and below the
transition temperature for Bose-Einstein condensation \cite{[17]}. These
studies on EIT and the related phenomena in the cold atoms provided
intensive understanding of the atomic coherence and interference in the
fundamental interaction between the light field and the atoms \cite%
{[18],[19],[20],[21],[22],[23]}.

EIT in the simple three-level system have been extensively studied, but EIT
in the multi-level cascade systems and their possible applications have not
been fully explored. Although essential physics about EIT has been
understood well from the studies of the simple three-level systems, there
are several interesting features in the complicated multi-level systems.
First, it is possible to create multiple EIT windows in such systems, which
will simultaneously support slow group velocities for two or more probe
pulses at different frequencies. As discussed in ref \cite{[24]}, two light
fields propagating with slow group velocities have advantage to efficiently
produce the quantum entanglement. The complicated multi-level systems
supporting slow group velocities for multiple light fields may provide such
interesting possibilities. Second, multi-level EIT systems may be useful for
nonlinear light generation processes \cite{[19],[25]}. In the simple
three-level EIT system, degenerate four-wave mixing can be enhanced due to
the reduction of the light absorption near the EIT window. Since the
complicated multi-level systems exhibit multiple EIT windows, the
non-degenerate four-wave mixing processes may be efficiently produced by
choosing the pump fields with frequencies near the multiple EIT windows.

Recently, McGloin et al \cite{[26]} theoretically studied an N-level EIT
system with several coupling fields and found that the spectral profiles
were quite different from that in the simple three-level EIT systems. The
cascade EIT system was studied in a vapor cell at room temperatures \cite%
{[27],[28]} and the EIT for resonant two-photon transitions in rubidium
atomic vapors was reported by Xu et al \cite{[29]}. In our previous work,
the multi-window EIT produced by bichromatic coupling fields was
demonstrated \cite{[30]}. Here, we report an experimental study of EIT in a
multi-level cascade system in cold $^{85}$Rb atoms. We apply an intense
coupling field that is nearly resonant with the three sets of transitions
among the $5P_{3/2}$and $5D_{5/2}$ levels. We investigate the probe
absorption spectra with multiple spectral peaks and the dependence of the
probe absorption profiles on the intensities of the coupling laser. Our
experimental measurements agree with the theoretical calculations.

We consider cascade EIT system shown in Fig. 1(a), an intense coupling laser
with frequency $\omega _{c}$ drives the transition $|2\rangle \rightarrow
|3\rangle $, and a weak probe laser with frequency $\omega _{p}$ drives the
transition$|1\rangle \rightarrow |2\rangle $. $\Delta _{c}$ and $\Delta _{p}$
are frequency detuning of the coupling laser and the probe laser,
respectively. EIT occurs when the coupling and the probe laser frequency
satisfy the two-photon resonant condition: $\Delta _{c}=\Delta _{p}$.
Scanning the probe laser around the transition $|1\rangle \rightarrow
|2\rangle $, a typical double-peaked EIT spectrum will appear as shown in
Ref. \cite{[31]}. However, for $^{85}$Rb atoms, the upper excited state $%
|3>(5D_{5/2})$ consists of three closely-spaced hyperfine levels $F^{\prime
\prime }=2,3,$ and 4 as shown in Fig. 1(b). The coupling laser will directly
drive three sets of transitions $5P_{3/2},F^{%
{\acute{}}%
}=3\rightarrow 5D_{5/2},F^{\prime \prime }=4,5P_{3/2},F^{%
{\acute{}}%
}=3\rightarrow 5D_{5/2},F^{\prime \prime }=3$, and $5P_{3/2},F^{%
{\acute{}}%
}=3\rightarrow 5D_{5/2},F^{\prime \prime }=2$. We define the coupling and
probe laser detuning as $\Delta _{c}$=$\omega _{c}$-$\omega _{0}$ and $%
\Delta _{p}$=$\omega _{p}$-$\omega _{21}$ respectively ($\omega _{0}$ is the
resonant frequency of the transition 5P$_{3/2}$, $F^{%
{\acute{}}%
}=3\rightarrow 5D_{5/2},F^{\prime \prime }=3$, $\omega _{21}$ is the
resonant frequency of the transition $5S_{1/2},F=3\rightarrow 5P_{3/2},F^{%
{\acute{}}%
}=3$). Because the frequency separations of the hyperfine components of $%
5D_{5/2}$ are quite small (less than 10 MHz \cite{[32]}), the $^{85}$Rb
cascade system will exhibit clear multi-window EIT effect even with a
moderately intense coupling laser. The probe absorption spectrum will
display a line profile with multiple peaks corresponding to the multiple
dressed states generated by the coupling between the three sets of
transitions, and the multiple transparent windows with minimum absorption
locate between the peaks. The corresponding probe dispersion will exhibit a
line profile with normal steep slopes at several frequencies near the
multiple transparent windows. Therefore, the multi-level cascade system
supports slow group velocities simultaneously for light pulses at different
frequencies.

We carry out a theoretical calculation of a five-level system as shown in
Fig.1 (a). The Rabi frequencies for the probe and coupling laser are defined
as $2\Omega _{p}=2\mu _{21}E_{p}/\hbar $ and $2\Omega _{c}=2\mu
_{32}E_{c}/\hbar $. The master equation is derived under the dipole
interaction and the rotating-wave approximation as

\begin{equation}
\dot{\rho}=-\frac{i}{\hbar }[H,\rho ]+\gamma _{21}L_{21}\rho +\gamma
_{32}L_{32}\rho +\gamma _{42}L_{42}\rho +\gamma _{52}L_{52}\rho   \label{1}
\end{equation}

Where the system Hamiltonian is written in the form

\begin{equation}
H=H_{0}+H_{I}  \label{2}
\end{equation}

$H_{\mathit{0}}$ is the free Hamiltonian and $H_{\mathit{I}}$ is the
interaction Hamiltonian,

\begin{equation}
H_{0}=-\hbar \Delta _{\text{p}}\sigma _{22}-\hbar (\Delta _{\text{p}}+\Delta
_{\text{c}}+\delta _{1})\sigma _{44}-\hbar (\Delta _{\text{p}}+\Delta _{%
\text{c}})\sigma _{33}-\hbar (\Delta _{\text{p}}+\Delta _{\text{c}}-\delta
_{2})\sigma _{55}  \label{3}
\end{equation}

\begin{equation}
H_{I}=-\hbar \Omega _{\text{p}}\sigma _{21}-\hbar a_{32}\Omega _{\text{c}%
}\sigma _{32}-\hbar a_{42}\Omega _{\text{c}}\sigma _{42}-\hbar a_{52}\Omega
_{\text{c}}\sigma _{52}+H.c.  \label{4}
\end{equation}

with $\delta _{1}=\omega _{34}=9$ MHz, $\delta _{2}=\omega _{53}=$ 7.6 MHz. $%
\sigma _{\mathit{ij}}=|i><j|$ ($i,j=1\sim 5$) are the population operators
when $i=j$ and dipole operators when $i\neq j$. $a_{32}$ , $a_{42}$ and $%
a_{52}$ are the relative strengths of the three transitions from the three
hyperfine sublevels $|3>$ , $|4>$ and $|5>$ ($F^{{\large {"}}}=3,4,2$) to
the state $|2>(F^{%
{\acute{}}%
}=3)$, and $a_{32}:a_{42}:a_{52}\approx 1:1.46:0.6.\gamma _{\mathit{ij}}$
denotes the spontaneous emission rates from level $i$ to level $j$, here $%
\gamma _{32}=\gamma _{42}=\gamma _{52}=\gamma =0.97$ MHz, $L_{\mathit{ij}%
}\rho $ describes the atomic decay from level $i$ to level $j$ and takes the
form

\begin{equation}
L_{ij}\rho=\frac{1}{2}(2\sigma_{ji}\rho\sigma_{ij}-\sigma_{ij}\sigma_{ji}%
\rho-\rho\sigma_{ij}\sigma_{ji})  \label{5}
\end{equation}

We numerically solve the density matrix equations of the five-level system
in the steady state. The calculated probe absorption [Im ($\rho _{21}$)] and
probe dispersion [Re($\rho _{21}$)] versus the probe detuning $\Delta _{p}$
are plotted in Fig. 2. The absorption spectrum exhibits three dips, and the
corresponding dispersion profile exhibits three steep normal slopes, which
should be useful for supporting slow light pulses at different frequencies.
Fig. 2(a), (b) and (c) show the calculated probe absorption and dispersion
profiles for the coupling Rabi frequencies $\Omega _{c}=4\gamma $, $7\gamma $%
, and 12$\gamma $ respectively, where the coupling detuning $\Delta
_{c}=-9\gamma $. We find that the frequency separations between the
neighboring absorption peaks are proportional to the coupling intensity, the
width and the location of the EIT windows can be controlled by the intensity
and the frequency detuning of the coupling laser. As one example, the
multi-level EIT system may be used to produce simultaneously two slow
photons at different frequencies, which may be used to generate photon
entanglement effectively \cite{[24]}.

Our experiment is carried out in a rubidium atom MOT \cite{[33]}, and the
experimental setup is briefly shown in Fig.3. The cooling and trapping
beam(780 nm) is supplied by a 500 mW diode laser system(TOPTICA TA100), its
frequency is stabilized by the saturated absorption spectra method, and its
linewidth is less than 1 MHz. The frequency of the trapping laser is
red-detuned about 12 MHz to the $F=3\rightarrow F^{\prime }=4$ transition by
using an acousto-optic modulator(AOM). The repumping laser (780 nm) from a
50 mW diode laser(TOPTICA DL100) is tuned to $F=2\rightarrow F^{\prime }$
transition of $^{85}$Rb $D_{2}$ line. We get a near spherical atom cloud
with the diameter of 3 mm in the MOT, containing about $5\times 10^{7}$
atoms, and the temperature of atom cloud is about 100 $\mu $K. The coupling
beam for EIT is taken from a Ti: sapphire laser (Coherent MBR110) and the
beam diameter is 3 mm. The transverse modes of the above lasers are filtered
to the fundamental Gaussian mode via single mode polarization maintenance
fibers. Another 50 mW diode laser (TOPTICA DL100) supplies weak probe beam,
the laser beam diameter is 1 mm and laser power is 1 $\mu $W. Both the
coupling and the probe beams are linearly polarized. They pass through the
atom cloud in opposite directions and overlap in the path via polarization
beam splitter (PBS). A photodiode (PD) is used to detect the probe light,
and a digital oscilloscope (Tektronix TDS 220) is adopted to monitor and
record the probe signals.

In order to avoid the influence of the MOT laser, we design a time sequence
using AOMs and functional generators (SRS DS345). The experiment is running
at a repetition of 10 Hz. In every 100 ms period, the cooling and trapping
process last for 97.5 ms, and the scanning of the probe absorption remains
2.5 ms. We find that there is no obvious difference in the absorption
spectra between the situations of MOT on and MOT off, so in most time our
experiment is carried out with the MOT on.

We tune the coupling laser nearly resonant to the transition $%
5P_{3/2},F^{\prime }=3\rightarrow 5D_{5/2},F^{\prime \prime }=2$, and the
probe laser is scanned across the transition $5S_{1/2},F=3\rightarrow
5P_{3/2},F^{\prime }=2,3,4$. When the coupling beam is blocked, we get
absorption spectrum of cold $^{85}$Rb atoms as shown by dashed line in Fig.
4, and the linewidth of each absorption peak is broader than the natural
linewidth of the corresponding transition. This is caused mainly by Zeeman
shift because we do not shut off the MOT magnetic filed during the period of
data acquiring. The solid line in Fig. 4 shows the absorption spectrum of
the probe laser with coupling laser on, the absorption peak is broadened and
is split into several peaks, and the dips between the absorption peaks are
probe transparencies induced by the coupling laser field. The system is
quite different from that described in Ref. [24], here we select $%
F=3\rightarrow F^{\prime }=3$ as the probe transition, and the possible
coupling transition for observation of the cascade EIT are $F^{\prime
}=3\rightarrow F^{\prime \prime }=2,3$ and 4. The space between the
sublevels $F^{\prime \prime }=2$ and $F^{\prime \prime }=3$ is 7.6 MHz, and
the space between sublevels $F^{\prime \prime }=3$ and $F^{\prime \prime }=4$
is 9.0 MHz. When the coupling laser is near resonance with the $F^{\prime
}=3\rightarrow F^{\prime \prime }=3$ transition, because of the small spaces
of $5D_{5/2}$ sublevels, one coupling laser tuned to $5P_{3/2},F^{\prime
}=3\rightarrow 5D_{5/2},F^{\prime \prime }$ transition can be regarded as
several coupling fields, but with different detuning for different $%
F^{\prime \prime }$. The reason of choosing such a level configuration is
that with the smaller separation of sublevels of $5D_{5/2}$ and the smaller
Zeeman shift of $5P_{3/2},F=2$ sublevel, we expect to obtain more EIT dips
with high resolution in cold atoms.

We tune coupling laser nearly resonant to the transition $5P_{3/2},F^{\prime
}=3\rightarrow 5D_{5/2},F^{\prime \prime }=3$, and investigate the
dependence of probe spectra on the coupling intensities. The experimental
results with $\Delta _{c}=-2$ MHz are shown in Fig. 5. Figure 5(a) is the
probe spectrum with the coupling Rabi frequency $\Omega _{c}=2$ MHz, the
solid curve is the experimental data, and the dashed line is the calculated
results with the spectral broadened arbitrarily to fit the experimental
results. Figures 5(b), 5(c), and 5(d) are for $\Omega _{c}=4,6$, and 9 MHz
respectively. With the higher coupling intensity, we obtain the deeper and
the broader transparency dips.

We perform an experimental study of multi-level cascade type
elctromagnetivcally induced transparency in cold $^{85}$Rb atoms confined in
a MOT. We obtain the unusual absorption profiles with multi-windows EIT. It
is possible to realize slow photons with multiple different frequencies
synchronously in such a multi-level cascade EIT system and to improve the
controllability of quantum information based on EIT. We investigate the
dependence of the probe absorption spectra on the coupling intensity. The
calculations based on density matrix equations agree with the experimental
results.

Support from the National Natural Science Foundation of China under Grant
Nos. 10104018 and 10374120 is acknowledged. YZ acknowledges support from the
National Science Foundation (Grant No. 001432).

$\bigskip $\linebreak 

$\mathbf{Figurecaptions}$

Fig. 1 Level configuration of the cascade type electromagnetically induced
transparency, (a) is schematic diagram; (b) is energy levels of cold $^{85}$%
Rb atoms.

Fig.2 The calculated probe absorption spectra of multi-level cascade EIT in
cold $^{85}$Rb atoms. The probe laser scans across the transition $%
5S_{1/2},F=3\rightarrow 5P_{3/2},F^{\prime }=3,\Delta _{c}=-9\gamma (\gamma
=0.97$ MHz). (a), (b) and (c) are for $\Omega _{c}=4\gamma ,7\gamma $, and $%
12\gamma $ respectively. The dashed lines are dispersion profiles.

Fig. 3 The experimental setup for electromagnetically induced transparency
in cold atoms confined in a magneto-optical trap. PBS: polarizing beam
splitter; PD: photodiode; L: lens; BS: beam splitter; M: mirror; AOM:
acousto-optic modulator.

Fig. 4 Absorption spectra of the multi-level cascade type EIT in cold $^{85}$%
Rb atoms with $\Omega _{c}=4$ MHz and $\Delta _{c}=-2$ MHz, the probe laser
scans across the transitions $5S_{1/2},F=3\rightarrow 5P_{3/2},F^{\prime
}=2,3,4$. The dashed line is the absorption spectrum without the coupling
laser, and the solid line is that with the coupling laser on. The linewidth
of the absorption peak is broadened by Zeeman shifts.

Fig. 5 The multi-level cascade type EIT spectra of $^{85}$Rb with the
different coupling laser intensities with $\Delta _{c}=-2$ MHz, the probe
laser scans across the transitions $5S_{1/2},F=3\rightarrow
5P_{3/2},F^{\prime }=3$. (a), (b), (c), and (d) are $\Omega _{c}=2,4,6$, and
9 MHz respectively. The solid curves are the experimental data and the
dashed lines are the calculated results with the spectral broadened
arbitrarily to fit the experimental linewidth.{\normalsize \newpage }

\end{document}